# Mapping the Empirical Evidence of the GDPR's (In-)Effectiveness: A Systematic Review

Wenlong Li[*], Zihao Li[†], Wenkai Li[‡], Yueming Zhang[§], Aolan Li[**]

**Abstract**

In the realm of data protection, a striking disconnect prevails between traditional domains of doctrinal, legal, theoretical, and policy-based inquiries and a burgeoning body of empirical evidence. Much of the scholarly and regulatory discourse remains entrenched in abstract legal principles or normative frameworks, leaving the empirical landscape uncharted or minimally engaged. Since the birth of EU data protection law, a modest body of empirical evidence has been generated but remains widely scattered and unexamined. Such evidence offers vital insights into the perception, impact, clarity, and effects of data protection measures but languishes on the periphery, inadequately integrated into the broader conversation. To make a meaningful connection, we conduct a comprehensive review and synthesis of empirical research spanning nearly three decades (1995- March 2022), advocating for a more robust integration of empirical evidence into the evaluation and review of the GDPR while laying a methodological foundation for coordinated research. By categorising evidence into four distinct groups– perception, effect, impact, and clarity, we provide a structured analysis therein and highlight the variety and nuances of the empirical evidence produced about the GDPR. Our discussion offers critical reflections on the current orientations and designs of evaluation work, challenging some popular but misguided orientations that significantly influence public debate and even direction of empirical and doctrinal research. This synthesis further sheds light on several understated aspects, surfaced by our systematic review, including the complex structure of the GDPR and the internal contradictions between components, the GDPR's interaction with other normative values and legal frameworks, as well as unintended consequences imposed by the GDPR on other values not explicitly recognised as regulatory objectives (such as innovation). We also propose a methodological improvement in how empirical evidence can be generated and utilised, underscoring the pressing need for more guided, coordinated and rigorous empirical research. By re-aligning empirical focus towards these ends and establishing strategic coordination at the community level, we seek to inform and underpin evaluative work that aligns empirical inquiries with policy and doctrinal

[*] Edinburgh Centre for Data, Culture & Society (CDCS), University of Edinburgh, UK.

[†] CREATe Centre, School of Law, University of Glasgow, UK; Stanford Law School, Stanford University, US. **Corresponding author** (Email: zihao.li@glasgow.ac.uk)

[‡] the Law, Science, Technology and Society Research Group (LSTS) and the Health and Aging Law Lab (HALL) of Vrije Universiteit Brussel, Belgium.

[§] Law & Technology research group, Ghent University, Belgium.

[**] Centre for Commercial Law Studies, School of Law, Queen Mary University of London

needs, while truly reflecting the complexities and challenges of safeguarding personal data in the digital age.

# I. Introduction

At the five-year enforcement milestone, the efficacy of the EU's General Data Protection Regulation (GDPR) in realising its articulated and anticipated objectives, or more broadly, in genuinely safeguarding personal data against a spectrum of recognised and potential threats to individual freedoms and fundamental rights, has increasingly come under scrutiny. The evaluation of the GDPR's effectiveness is undertaken at various fora and against varying criteria, ranging from the formal periodical evaluation by the European Commission as stipulated by the GDPR itself,[6] to the empirical probes conducted in academic scholarship,[7] by civil societies[8] or media.[9] From these multifarious evaluative contexts, a spectrum of arguments emerges, articulating both the intended and the observed effects of the GDPR, often manifesting with inconsistent or even conflicting perspectives, highlighting a broad array of interpretations regarding the regulation's impact. For instance, the European Commission's official review after two years of GDPR enforcement acknowledges the regulation's initial successes and delineates areas requiring enhancement. This recognition contrasts sharply with vigorous debates from civil society and academia, which critique the GDPR's enforcement deficits, lack of clear guidance, and operational challenges that impede the realisation of its full potential.[10] Commonly raised concerns include the efficacy of key GDPR components like the validity of consent and the exercisability of data subject rights. Additionally, attention is drawn to broader institutional, operational, and structural issues that demand a thorough empirical investigation.[11] Advocates from civil society, such as Brave[12], highlight the ineffectiveness of enforcement against major technology firms, attributing this to the inadequate resources of competent authorities, among other factors. Industry-supported research argues that the GDPR stifles technological innovation, with claims that the law is 'killing innovative apps'.[13] These diverse viewpoints have resonated in recent policy discourses. For example, the UK's post-Brexit reform of its Data Protection Act advocates for an 'innovative approach', branding it as a 'new direction' for data protection that emphasises the law's role in fostering digital innovation.[14] Similarly, in the EU, efforts to streamline GDPR procedural aspects have led to the rapid introduction of accompanying procedural regulations, reflecting a proactive approach to refining the regulatory framework.[15]

---

[6] European Commission, 'Commission Report: EU Data Protection Rules Empower Citizens and Are Fit for the Digital Age' (2020) <https://ec.europa.eu/commission/presscorner/detail/en/ip_20_1163>.

[7] Tuulia Karjalainen, 'The Battle of Power: Enforcing Data Protection Law against Companies Holding Data Power' (2022) 47 Computer Law & Security Review 105742 <https://linkinghub.elsevier.com/retrieve/pii/S0267364922000851>; Filippo Lancieri, 'Narrowing Data Protection's Enforcement Gap' (2022) 74 Maine Law Review, <https://www.ssrn.com/abstract=3806880>.

[8] Johnny Ryan, 'Europe's Governments Are Failing the GDPR: Brave's 2020 Report on the Enforcement Capacity of Data Protection Authorities' (2020).

[9] Matt Burgess, 'How GDPR Is Failing' (*WIRED*, 2022) <https://www.wired.co.uk/article/gdpr-2022> accessed 27 May 2023.

[10] David Erdos, 'Acontextual and Ineffective? Reviewing the GDPR Two Years On' (*International Forum for Responsible Media*, 2020) <https://inforrm.org/2020/05/05/acontextual-and-ineffective-reviewing-the-gdpr-two-years-on-david-erdos/>; Ryan (n 8).

[11] Ilse Heine, '3 Years Later: An Analysis of GDPR Enforcement' (*Centre for Strategic & International Studies (CSIS)*, 2021); Erdos (n 10); A Fluitt and others, 'Data Protection's Composition Problem' (2019) 5 European Data Protection Law Review 285 <http://edpl.lexxion.eu/article/EDPL/2019/3/4>; Salomé Viljoen, 'A Relational Theory of Data Governance' (2021) 131 Yale Law Journal 370; Karen Yeung and Lee A Bygrave, 'Demystifying the Modernized European Data Protection Regime: Cross‐disciplinary Insights from Legal and Regulatory Governance Scholarship' (2022) 16 Regulation & Governance 137 <https://onlinelibrary.wiley.com/doi/10.1111/rego.12401>.

[12] Ryan (n 8).

[13] Rebecca Janßen and others, 'GDPR and the Lost Generation of Innovative Apps' [2022] National Bureau of Economic Research <https://www.nber.org/papers/w30028#:~:text=Using data on 4.1 million,new apps fell by half>.

The degree to which reforms and evaluative discussions around the GDPR are grounded in empirical evidence remains a contentious issue. Although surveys and statistical data are routinely presented at various levels and by different stakeholders, as part of the legislative review or reform processes, these metrics are typically partial and/or perceptive, hence providing limited insights into the actual functionality and tangible impacts of GDPR mechanisms. For instance, the Centre for European Policy Studies report, in its evaluation of the GDPR's effectiveness, uses 'two case studies' (respectively on the privacy shield and the Convention 108+) to provide a snapshot of how the GDPR is operated.[16] In a widely distributed op-ed on WIRED, Burgess makes a powerful case for 'the failing GDPR',[17] pointing to the figures on the number of cases resolved and the volume of fines imposed. Even the legally mandated review after two years of GDPR implementation, the European Commission claimed 'overall success' on the basis of the survey conducted by the EU Fundamental Rights Agency about public awareness of the GDPR, together with a set of numbers indicative of the actions taken by authorities at both EU and Member States levels.[18]

There is a discernible underutilisation or recognition of the rich empirical evidence that has accumulated within the legal scholarship. This can be attributed not only to a historical lack of synthesis over the last two decades but also the insufficient coordination and direction in empirical research efforts arising from different domains. Such underestimation of evidence may might have been justified in the early years of GDPR enforcement when the components of EU data protection law operated primarily at the concept or principle level. This is, however, no longer the case, as a spontaneous and often disparate surge is seen in empirical efforts across various disciplines, with scholars and practitioners independently exploring the regulation's impact since the GDPR's implementation. Computer scientists tend to focus on operationalising high-level principles such as data minimization and purpose limitation, which lack detailed, actionable provisions under the GDPR.[19] Human-Computer Interaction (HCI) researchers have significantly contributed from a design perspective, particularly embracing the officially recognized principle of privacy by design.[20] Concurrently, there is a pronounced initiative within the IT and other industries to assess the practical implications of the law for specific sectors, groups, or practices.[21] Investigative journalists and civil society organizations have also generated evidence reflecting the under-enforcement against 'big tech.'[22] Additionally, diverse efforts across fields have aimed to capture the perceptions of various practitioner groups or the general public regarding the GDPR's effectiveness, interpreting these as compliance readiness, maturity, awareness, the feeling of being protected, or the adaptability of compliance mechanisms to the new law.[23] These varied and multidisciplinary empirical undertakings significantly enrich our understanding of the GDPR's effects and effectiveness. However, the spontaneous and uncoordinated nature of these efforts underscores an urgent need for a unified narrative and systematic coordination. This fragmented landscape of empirical research, starkly contrasting with the more extensive doctrinal work, reveals the relative unfamiliarity of the data protection community with empirical findings and approaches. Such a disconnect, as we will further argue, holds significant implications for evaluative endeavours, doctrinal development, and, more broadly, the understanding of personal data protection within EU contexts and beyond.

---

[14] UK Department for Digital Culture Media & Sport, 'Data: A New Direction - Government Response to Consultation' (2022) <https://www.gov.uk/government/consultations/data-a-new-direction/outcome/data-a-new-direction-government-response-to-consultation>.

[15] For example: Proposal for a REGULATION OF THE EUROPEAN PARLIAMENT AND OF THE COUNCIL laying down additional procedural rules relating to the enforcement of Regulation (EU) 2016/679

[16] Aengus Collins, 'A Recipe For Success? Assessing the EU's Actorness and Effectiveness in the Data Protection Domain' (2023) <https://cdn.ceps.eu/wp-content/uploads/2023/04/CEPS-2023-09-In-depth-analysis_EU-Data-Protection.pdf>.

[17] Burgess (n 9).

[18] European Commission (n 6).

[19] Asia J Biega and others, 'Operationalizing the Legal Principle of Data Minimization for Personalization', *Proceedings of the 43rd International ACM SIGIR Conference on Research and Development in Information Retrieval* (ACM 2020) <https://dl.acm.org/doi/10.1145/3397271.3401034>.

[20] Christine Utz and others, '(Un)Informed Consent: Studying GDPR Consent Notices in the Field', *Proceedings of the 2019 ACM SIGSAC Conference on Computer and Communications Security* (ACM 2019) <https://dl.acm.org/doi/10.1145/3319535.3354212>; Midas Nouwens and others, 'Dark Patterns after the GDPR:

The GDPR has swiftly emerged as a focal point for empirical analysis with an accumulating body of evidence about this perception, enforcement and broader implications rapidly converging. As we observe, research on EU data protection law is undergoing an "empirical turn" of its own, with a substantial increase in empirical research since the final approval of the GDPR in 2016, and this trend continues to grow.[24] However, visibility of this empirical research within the community

---

Scraping Consent Pop-Ups and Demonstrating Their Influence', *Proceedings of the 2020 CHI Conference on Human Factors in Computing Systems* (ACM 2020) <https://dl.acm.org/doi/10.1145/3313831.3376321>; Dominique Machuletz and Rainer Böhme, 'Multiple Purposes, Multiple Problems: A User Study of Consent Dialogs after GDPR' (2020) 2020 Proceedings on Privacy Enhancing Technologies 481 <https://petsymposium.org/popets/2020/popets-2020-0037.php>; Konrad Kollnig and others, 'A Fait Accompli? An Empirical Study into the Absence of Consent to Third-Party Tracking in Android Apps', *SOUPS'21: Proceedings of the Seventeenth USENIX Conference on Usable Privacy and Security* (USENIX Association 2021) <https://dl.acm.org/doi/abs/10.5555/3563572.3563582>.

[21] Dimitra Kamarinou, Christopher Millard and W Kuan Hon, 'Cloud Privacy: An Empirical Study of 20 Cloud Providers' Terms and Privacy Policies—Part I' (2016) 6 International Data Privacy Law 79 <https://doi.org/10.1093/idpl/ipw003>; Dimitra Kamarinou, Christopher Millard and W Kuan Hon, 'Cloud Privacy: An Empirical Study of 20 Cloud Providers' Terms and Privacy Policies—Part II' (2016) 6 International Data Privacy Law 170 <https://academic.oup.com/idpl/article-lookup/doi/10.1093/idpl/ipw004>; Lisa Parker and others, 'How Private Is Your Mental Health App Data? An Empirical Study of Mental Health App Privacy Policies and Practices' (2019) 64 International Journal of Law and Psychiatry 198 <https://www.sciencedirect.com/science/article/pii/S0160252718302681>; T Mulder and M Tudorica, 'Privacy Policies, Cross-Border Health Data and the GDPR' (2019) 28 Information and Communications Technology Law 261 <https://doi.org/10.1080/13600834.2019.1644068>; Trix Mulder, 'Health Apps, Their Privacy Policies and the GDPR' (2019) 10 European Journal of Law and Technology 1 <https://papers.ssrn.com/abstract=3506805>; Guy Aridor, Yeon-Koo Che and Tobias Salz, 'The Effect of Privacy Regulation on the Data Industry: Empirical Evidence from GDPR' (2023) 54 The RAND Journal of Economics 695 <http://www.nber.org/papers/w26900.pdf>.

[22] Burgess (n 9).

[23] For example, Wanda Presthus and Hanne SØrum, 'Are Consumers Concerned about Privacy? An Online Survey Emphasizing the General Data Protection Regulation' (2018) 138 Procedia Computer Science 603 <https://doi.org/10.1016/j.procs.2018.10.081>; Bruno Škrinjarić, Jelena Budak and Edo Rajh, 'Perceived Quality of Privacy Protection Regulations and Online Privacy Concern' (2019) 32 Economic Research-Ekonomska Istraživanja 982 <https://doi.org/10.1080/1331677X.2019.1585272>; Mirkó Gáti and AE Simay, 'Perception of Privacy in the Light of GDPR' [2020] 11th Proceedings of the European Marketing Academy; Mulder and Tudorica (n 21); Joanna Strycharz, Jef Ausloos and Natali Helberger, 'Data Protection or Data Frustration? Individual Perceptions and Attitudes towards the GDPR' (2020) 6 European Data Protection Law Review 407.

[24] See figure 1

remains limited due to the absence of meaningful and timely integration. To our best knowledge, despite the substantial passage of time since the inception of the EU data protection directive in 1995, there has been no comprehensive synthesis or integration of the empirical work produced on data protection law. Consequently, little is known within the academic community about the empirical landscape of the GDPR or its multiple and heterogeneous components, let alone using this evidence to inform, challenge, and refine existing theoretical debates or policymaking.

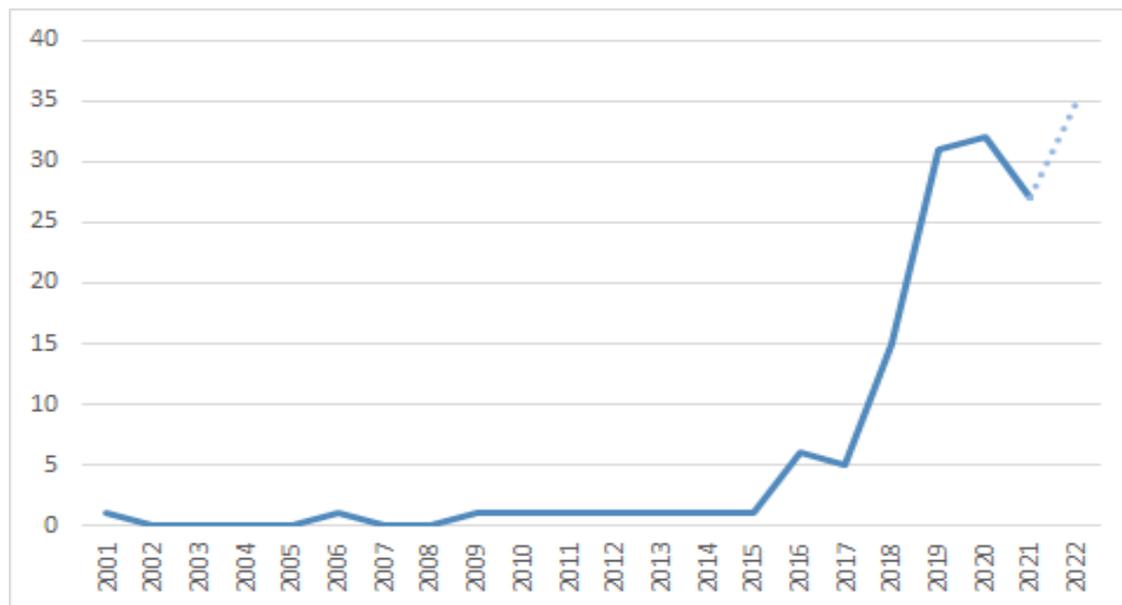

Figure 1 Distribution of empirical articles over time

Overall, the matter of effectiveness transcends mere compliance metrics, reaching into the realms of substantive data protection, which may be variably interpreted as individual autonomy over personal data, fair and accountable data-driven businesses, or good dealing with the societal implications of data processing activities. As such, effectiveness is not a monolithic construct but a blend of various dimensions that reflect the regulation's broad legislative intent and application across diverse contexts. The evaluation of effectiveness elucidates a profound disjunction between theoretical constructs and empirical validations, manifesting a critical need to reconcile diverse perspectives under a cohesive evaluative framework. It also prompts a re-evaluation of popular or conventional methods, which often fail to capture the multi-faceted impacts of the GDPR, skewing the regulatory narrative towards either partial criticism or overzealous acclaim. The challenge here lies in developing an integrative framework that not only respects the inherent complexities of the GDPR but also aligns with the empirical realities observed in its implementation.

The evaluation of the EU data protection law is essentially an empirical matter, yet the evidence produced in association with legislative review and reform is limited in nature, type and significance, and the body of empirical scholarship produced since the law's inception in 1995 is barely systematised. This paper intends to fill this gap by offering a systematic review of the empirical literature on the EU's Data Protection law spanning nearly three decades (1995 - March 2022). The material scope of our review covers not only the latest EU data protection law, i.e., the EU's GDPR but also its predecessor (the Data Protection Directive) and sibling instruments (i.e. the EU Data Protection Regulation (EUDPR) and the Law Enforcement Directive). We further reflect critically, based on our systematic review, how the effectiveness of the GDPR may be better evaluated in an evidence-based, rigorous and critical manner by identifying major challenges and providing actionable recommendations.

This paper proceeds in five parts. Following this introduction, Part two briefly outlines the methods employed for our systematic review. In Part Three, our synthesis of evidence is presented in four categories, including perception, effect, impact and clarity. In addressing the matter of GDPR's (in-)effectiveness, we offer our reflective and critical discussion on the nature, relevance and usefulness of the evidence reviewed in part Four. Part Five briefly concludes.

## II. Methods

We conducted the systematic review of the empirical research on EU data protection laws in accordance with the Preferred Reporting Items for Systematic reviews and Meta-Analyses (PRISMA) guidelines (see Figure 2).[25]

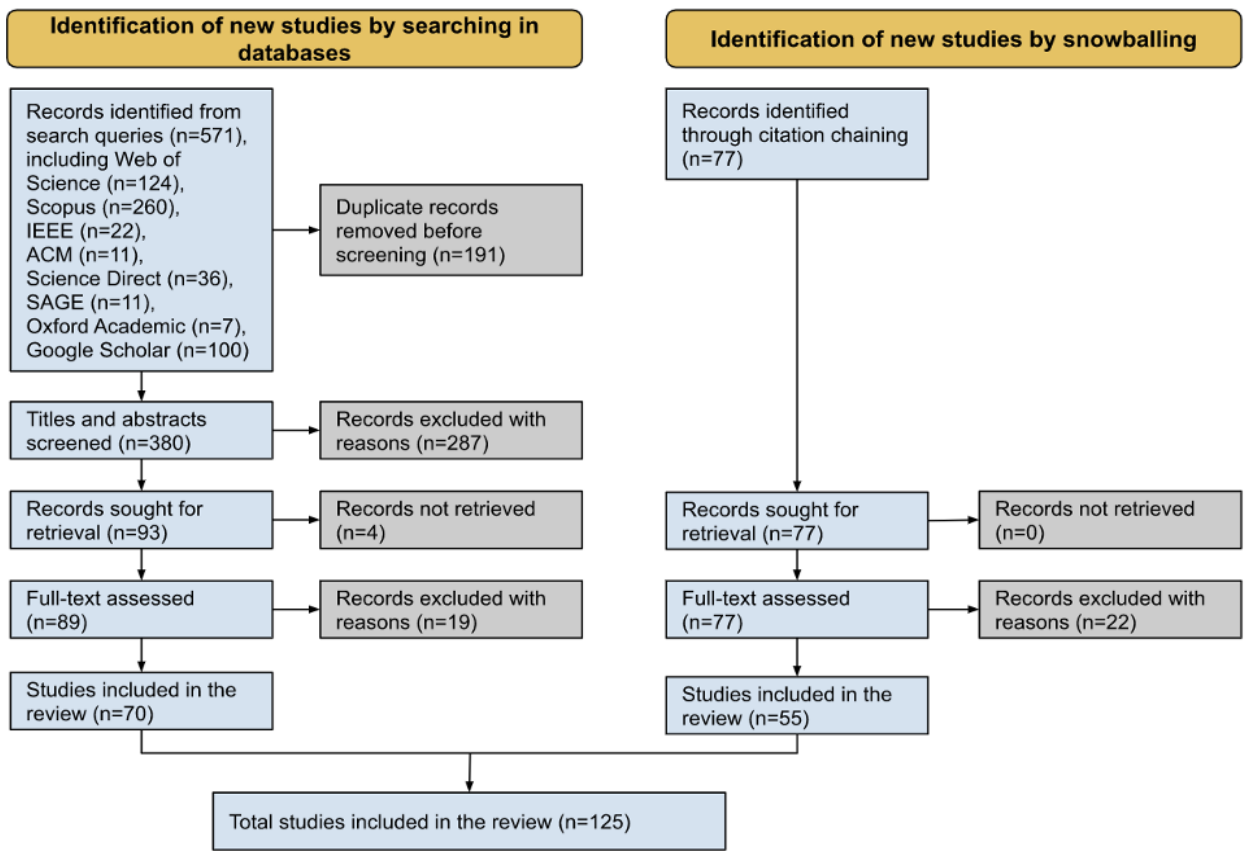

Figure 2 PRISMA flowchart

A combination of keywords – ('GDPR' OR 'data protection') AND empirical and its variations where search rules are different – is used to locate and identify relevant literature in several formal databases, including Web of Science, Scopus, IEEE, ACM, ScienceDirect, SAGE and Oxford Academic (see Figure 3). Google Scholar was also searched for double-check purposes and, given the high volume of results returned from Google (approx. 17, 100), we inspected the top 100 results marked as the most relevant. The following filters were deployed in our searches to narrow down the scope:

1. Timescale: published between 1995 and 2022 (March)
2. Language: written in English
3. Type: journal articles, conference papers, books and book chapters
4. Papers with preprints available online but were not officially published are excluded at the time of review

---

[25] Matthew J Page and others, 'The PRISMA 2020 Statement: An Updated Guideline for Reporting Systematic Reviews' [2021] The British Medical Journal 1 <https://www.bmj.com/lookup/doi/10.1136/bmj.n71>.

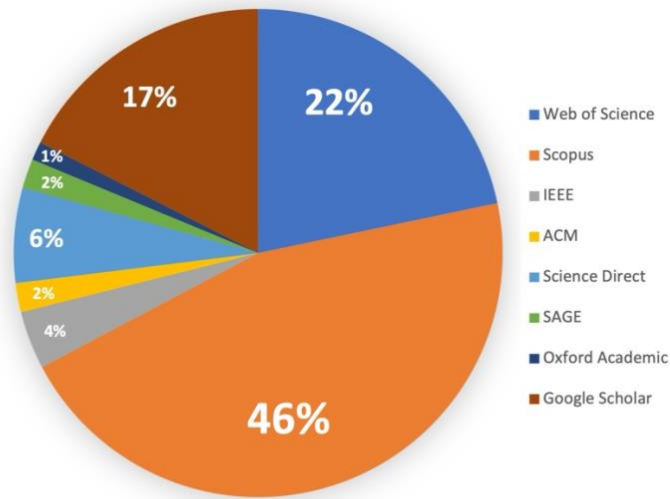

Figure 3 Distribution of included articles in our sample across 8 databases

Our searches across databases returned 380 papers in total from the outset, with some duplicates removed. Both title- and abstract-level screening were then conducted, in accordance with the following inclusion and exclusion criteria:

1. Relevance: the article should be explicitly about the EU Regulation 2016/679 (GDPR), Regulation 2018/1725 (EUDPR), Directive 2016/680 (the Law Enforcement Directive), or Directive 95/46/EC (Data Protection Directive)
2. Quality: the article should be peer-reviewed and of sound quality
3. Empirics: the article should include first-hand primary data or official secondary data, qualitative or quantitative, which are collected as part of the study and report.

Each paper was screened by at least two independent researchers who decided by voting. In case of disagreement, a third person intervened to provide a second opinion and final vote. Group discussion was convened for controversial entries whenever needed. In this process, we established further criteria to determine if a paper falls within the scope:

1. A paper is excluded if not explicitly grounded on one of the abovementioned EU legal instruments despite the existence of a general reference to the concept of privacy or data protection.
2. A paper is included if it engages the national implementation of the Directive 95/46/EC.
3. A paper is included if its methodology is defined as a 'case study' with qualitative or quantitative data collected and analysed.

## III. Findings

In this section, we present a qualitative synthesis of the empirical research on the GDPR, grouping them in four main clusters, including perception (how different stakeholders perceive the effect and compliance maturity of the GDPR), effect (the intended effect the law has on organizations and individuals), impact (its broader impacts on society and the economy), and clarity (the empirical work seeking to clarify or operationalise broad concepts and principles)

Due to the unsynchronised nature of the underlying studies, this synthesis does not attempt to establish demographic or statistical patterns but instead focuses on identifying thematic consistencies and discrepancies across the evidence collected. This approach highlights the diverse landscape of GDPR enforcement and provides a nuanced understanding of its multifaceted implications.

### 1. Perception

A major cluster of evidence produced about the EU data protection law is of a perceptive nature, inquiring actors of various kinds (including the general public, in-house lawyers and practitioners, as well as authorities and courts) of their overall impression, experiences and expected outcomes. Around the inception of the GDPR, questionnaire-based surveys report an increased level of public awareness.[26] This has occasionally been translated into a reduced concern of privacy or enhanced willingness to share personal information, sensitive information included, on the assumption that such data can be well taken care of and processed in a limited and expected manner.[27] Yet, counterevidence does exist to indicate distrust of the GDPR and its institutions as a genuine guardian of privacy and personal data.[28] Strycharz et al. point out, for instance, that a significant portion of the Dutch population depreciate the legislative efforts and cast doubts on the usefulness of data subject rights despite the high level of public awareness.[29]

More sophisticated inquiries extend beyond perception and probe its determining factors. For example, online survey by Pleger et al. of 1000 participants (50% UK and 50% Germany) show a general lack of adequate understanding of the terminologies (particularly the concept of data protection vis-à-vis data security) used in news articles and public debates.[30] With a quantitative survey of over 26000 European internet users, Prince et al. contest the claim that adoption of privacy-protective measures results in a reduction of privacy concerns.[31] As they claim, users with higher privacy literacy tend to have increased concerns about privacy, and that an increase in the knowledge of the data protection law does not lead to reduced concerns about their privacy. Similarly, empirical investigations into consumer inertia or the "privacy paradox"[32] conclude that improved awareness and confidence in the new legislation does not necessarily lead to action, since the altered social norms might also play a crucial role in promoting data sharing.[33] As privacy 'concernedness' is assessed in empirical research varies significantly in their scope/jurisdiction,[34] context (e.g., drones)[35] and targeted individuals (e.g., university students),[36] any attempt to generalise the level of awareness, understanding and preference as elevated by the enactment of a data privacy legislation are contested.

---

[26] Presthus and SØrum (n 23); Škrinjarić, Budak and Rajh (n 23); Gáti and Simay (n 23).
[27] M Karampela, S Ouhbi and M Isomursu, 'Exploring Users' Willingness to Share Their Health and Personal Data under the Prism of the New GDPR: Implications in Healthcare', *2019 41st Annual International Conference of the IEEE Engineering in Medicine and Biology Society (EMBC)* (IEEE 2019) <https://ieeexplore.ieee.org/document/8856550/>.
[28] Presthus and SØrum (n 23).
[29] Strycharz, Ausloos and Helberger (n 23).

From the perspectives of practitioners, broadly defined to include in-house lawyers, data protection officers and staff, privacy engineers etc., the perception of GDPR's effectiveness is often assessed in sector-specific contexts, e.g., IoT,[37] cloud computing,[38] tracking,[39] employer compliance[40] and the public sector.[41] This strand of empirical work features content analysis (e.g., inspection of privacy policies), with alternative methods such as controlled experiment.[42]

---

[30] Lyn E Pleger, Katharina Guirguis and Alexander Mertes, 'Making Public Concerns Tangible: An Empirical Study of German and UK Citizens' Perception of Data Protection and Data Security' (2021) 122 Computers in Human Behavior 106830 <https://doi.org/10.1016/j.chb.2021.106830>.

[31] Christine Prince and others, 'Are We Living in Surveillance Societies and Is Privacy an Illusion? An Empirical Study on Privacy Literacy and Privacy Concerns' [2021] IEEE Transactions on Engineering Management 1.

[32] Wanda Presthus and Hanne Sørum, 'Consumer Perspectives on Information Privacy Following the Implementation of the GDPR' (2019) 7 International Journal of Information Systems and Project Management 19; Presthus and SØrum (n 23).

[33] Jeff Jarvis, *Public Parts How Sharing in the Digital Age Improves the Way We Work and Live* (Simon & Schuster 2015); Strycharz, Ausloos and Helberger (n 23).

[34] Škrinjarić, Budak and Rajh (n 23).

[35] Rachel L Finn and David Wright, 'Privacy, Data Protection and Ethics for Civil Drone Practice: A Survey of Industry, Regulators and Civil Society Organisations' (2016) 32 Computer Law and Security Review 577 <http://dx.doi.org/10.1016/j.clsr.2016.05.010>.

[36] Gáti and Simay (n 23).

[37] Sarah Turner and others, 'The Exercisability of the Right to Data Portability in the Emerging Internet of Things (IoT) Environment' (2021) 23 New Media & Society 2861 <https://doi.org/10.1177/1461444820934033>.

[38] Dimitra Kamarinou, Christopher Millard and Isabella Oldani, 'Compliance as a Service' [2018] Queen Mary University of London Legal Studies Research 34 <https://papers.ssrn.com/sol3/papers.cfm?abstract_id=3284497%0Ahttp://www.mccrc.org/events/2018-symposium-compliance-as-a-service/>.

[39] Mulder and Tudorica (n 21); Mulder (n 21); Miguel Godinho de Matos and Idris Adjerid, 'Consumer Consent and Firm Targeting After GDPR: The Case of a Large Telecom Provider' (2022) 68 Management Science 3330 <http://pubsonline.informs.org/doi/10.1287/mnsc.2021.4054>.

[40] Seili Suder and Andra Siibak, 'Employers as Nightmare Readers: An Analysis of Ethical and Legal Concerns Regarding Employer-Employee Practices on SNS' (2017) 10 Baltic Journal of Law & Politics 76 <https://www.sciendo.com/article/10.1515/bjlp-2017-0013>.

[41] Dominika Lisiak-Felicka and Maciej Szmit, 'GDPR Implementation in Public Administrationin Poland – 1.5 Year after: An Empirical Analysis' (2021) 43 Journal of Economics and Management 1 <https://sbc.org.pl/Content/452500/01_02.pdf>.

[42] Godinho de Matos and Adjerid (n 39).

Controversially, the measurement of practitioners' perception about the GDPR often centres on privacy policies only. [43] For instance, Kamarinou et al. inspect data processing agreements of 13 cloud service providers and show updates in line with the GDPR, including provisions on the supply of tools and services intended to assist customers in their compliance.[44] By comparing three corpora of privacy policies (n=550, 450) before and after the GDPR, Zaeem and Barber show clear evidence of progress made at a textual level and identify several problematic areas (e.g., data subject rights, and data sharing with law enforcement) where the disclosure of information is terse, ambiguous or inadequate.[45] In a similar vein, Mulder and Tudorica's inspection of the data-driven wearable companies is marked with ambiguity of purposes for which data are processed and shared and with whom.[46] In collaboration with local rehabilitation centres in the Netherlands, Mulder identifies 31 wearable apps that process health data and compared their privacy policies against the requirements of the GDPR and concludes that, while the policies market with their claimed compliance with the GDPR, they are deliberately and unsatisfactorily vague, again, on the purpose for which the health data are processed. [47]

Empirical endeavours also extend beyond textual promises. For example, Godinho de Matos and Adjerid engage 33,629 client households of TELCO with a "A/B test", and their results show a counterintuitive case where the rate of customers giving consent increases for the use of household service usage (27.6% cf 4.3%), profiling (64% cf 53%), and even geolocation tracking and third-party tracking (4.7% cf 3.5%),[48] thus providing intriguing insights into customer behaviour in relation to consent under the GDPR. Regarding compliance in the public sector, Lisiak-Felicka & Szmit conduct web interviews with Polish public administration offices.[49] As they observe, the majority of districts (approx. 80%) and all marshal offices confirmed general compliance, but 23% of the municipal offices declared their inability to do the same due to the lack of training for employees, management support, limited budgets etc.[50] With surveys completed by 233 Czech organisations (response rate 14.87%), Faifr and Januška identify factors that determine the scope and "cost intensity" of GDPR implementation projects, notably including organisation size (regardless of public or private), data types and implementation strategies.[51]

The perspective of enforcement and interpretation by authorities is the least examined in this cluster, with the empirical focus placed upon the logic, landscape and predictability of imposing fines by authorities. Presthus and Sønslien, based on a sample of 277 sanctions documented in two popular GDPR databases (PRIVACY Affairs & GDPR Enforcement Tracker), conclude that the lack of an adequate legal basis for processing stands as the top reason for imposing administrative fines (followed by third-party access and tracking, data subject rights, and security), and that the volume of fines varies significantly between the symbolic 90 euros to 50 million.[52] Ruohonen and Hjerppe's work confirms the overall landscape of the enforceability of various GDPR provisions, as well as the great variation as shown by different authorities across the EU.[53] Mantelero and Esposito's inspection of 700 decisions and documents released by six European data protection authorities has a normative objective, i.e., to deduce patterns and strategies crucial to develop a methodological framework for the so-called 'Human Rights Impact Assessment' (HRIA).[54] Qualitative data are also gathered from authorities, for instance by Presthus and Sønslien[55], to reveal rationales for the content analysis of 277 sanctions they conducted beforehand. Ceross's semi-structured interviews with case officers from the ICO extend beyond publicly available information and shows that the ICO's resource constraint does have an impact on its enforcement priorities and its preference of cooperation over adversariality.[56] In our sample, there is only one paper collecting empirical data from the judiciary but mostly at the demographical levels, e.g. the percentage of data protection cases adjudicated by the Court of justice, demographic features of the parties involved.[57]

---

[43] Razieh Nokhbeh Zaeem and K Suzanne Barber, 'The Effect of the GDPR on Privacy Policies: Recent Progress and Future Promise' (2020) 12 ACM Transactions on Management Information Systems 2:1 <https://doi.org/10.1145/3389685>.
[44] Kamarinou, Millard and Oldani (n 38).
[45] Zaeem and Barber (n 43).
[46] Mulder and Tudorica (n 21).
[47] Mulder (n 21).
[48] Godinho de Matos and Adjerid (n 39).
[49] Lisiak-Felicka and Szmit (n 41).
[50] ibid.
[51] Adam Faifr and Martin Januška, 'Factors Determining the Extent of GDPR Implementation within Organizations: Empirical Evidence from Czech Republic' (2021) 22 Journal of Business Economics and Management 1124 <https://journals.vilniustech.lt/index.php/JBEM/article/view/15095>.

Administrative fines stand, understandably, as the most attended empirical issue. Whereas some compare the varying fines imposed on the same ground by different countries,[58] others leverage machine learning methods to draw patterns from judicial meta-data and textual traits with a view to achieve reliable prediction on the GDPR fines. Based on meta-data and textual traits, Ruohonen and Hjerppe inquire the extent to which the GDPR fines can be predicted using basic machine learning methods.[59] As they conclude, both meta-data as provided by private third-party websites and textual features extracted from official documents can be used (with the latter outperforming the former) to make reliable, black-box predictions with average error below ten euros.

## 2. Effect

There are little or no empirical efforts to evaluate the GDPR's effect with reference to its stated objectives, which is defined in this paper as the extent to which the GDPR is operated as expected. The measurement of effect is primarily undertaken at a granular level. The GDPR is indeed hybrid in nature, consisting of a variety of mechanisms drawn across several disciplines, between which there can be potential conflicts and tension as reflected in the doctrinal research.[60] Some empirical work engages a wider range of legal areas such as consumer protection and contract,[61] whereas others refer to the GDPR only as a buzzword. Empirical attention paid to the GDPR components are also highly uneven: as shown in the figure 4, the right of access stands as the most popular site of empirical inquiry), which has received more attention than all other subject rights combined. In contrast, there are few instances in which principles, data protection impact assessment, and security are subject to empirical examination. Several other mechanisms have, thus far, never been subject to any form of empirical consideration. In light of this highly uneven empirical landscape (largely determined by the ease of data collection and analysis), evidence provided in this section is admittedly partial, concerning primarily topical mechanisms such as privacy policies, consent, dark patterns, and data subject rights.

---

[52] Wanda Presthus and Kaja Felix Sønslien, 'An Analysis of Violations and Sanctions Following the GDPR' (2021) 9 International Journal of Information Systems and Project Management 38.
[53] Jukka Ruohonen and Kalle Hjerppe, 'The GDPR Enforcement Fines at Glance' (2022) 106 Information Systems 101876 <https://doi.org/10.1016/j.is.2021.101876>.
[54] Alessandro Mantelero and Maria Samantha Esposito, *An Evidence-Based Methodology for Human Rights Impact Assessment (HRIA) in the Development of AI Data-Intensive Systems*, vol 41 (Elsevier Ltd 2021) <https://doi.org/10.1016/j.clsr.2021.105561>.
[55] Presthus and Sønslien (n 52).
[56] Aaron Ceross, 'Examining Data Protection Enforcement Actions through Qualitative Interviews and Data Exploration' (2018) 32 International Review of Law, Computers and Technology 99 <https://doi.org/10.1080/13600869.2018.1418143>.
[57] Ondřej Pavelek and Drahomíra Zajíčková, 'Personal Data Protection in the Decision-Making of the CJEU before and after the Lisbon Treaty' (2021) 11 TalTech Journal of European Studies 167.
[58] Presthus and Sønslien (n 52).
[59] Ruohonen and Hjerppe (n 53).

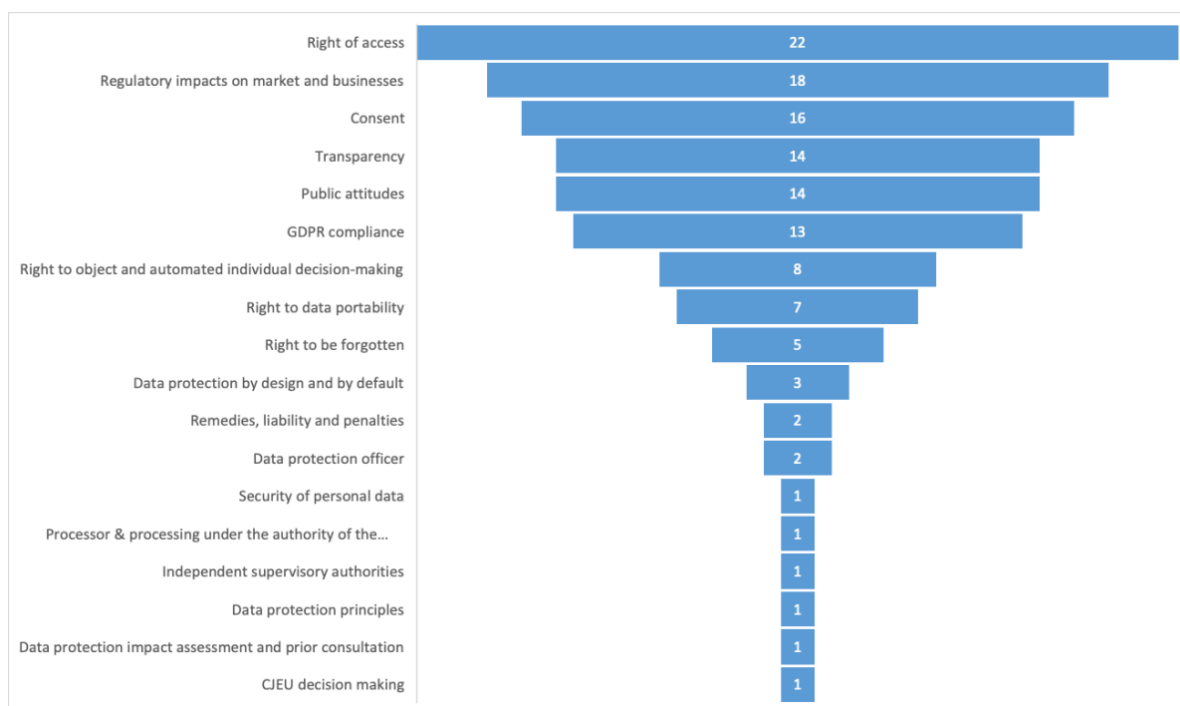

Figure 4 Statistics on the frequency of the GDPR components being empirically examined

Privacy policy has been a popular site of empirical inquiry for its ease of accessibility and regular archival. The impact of the data protection law on privacy policies has extensively examined within diverse contexts e.g., mental health apps[62] and public sector websites.[63] As a business practice emerged in the late 1990s, it predates the legal requirements for the transparency of personal data processing. The practice began to gain prominence in late 1990s and early 2000s when industry and advocacy groups sought to 'self-regulate' by adopting privacy policies as a means to address the concerns of privacy and data protection.[64] The self-regulatory nature of these practices has been however arguable as legal requirements of greater transparency started to mount at the same time, taking the EU's Data Protection Directive of 1995 as an example. Now, privacy policies are significantly shaped by transparency requirements of the law, with the GDPR consisting of granular and ultra-lengthy provisions on what information should be provided to the data subject and in what manner and additional transparency requirements to be introduced by the DMA and DSA in the EU. The date the GDPR came into effect is an important marker around which comparative or longitudinal research has been conducted.[65] With a new set of fine-grained, detailed requirements, the GDPR is expected to make a difference to how privacy policy is formulated in practice. However, recent empirical inspection strongly contests this assumption.

---

[60] Michael Veale, Reuben Binns and Jef Ausloos, 'When Data Protection by Design and Data Subject Rights Clash' (2018) 8 International Data Privacy Law 105 <https://academic.oup.com/idpl/article/8/2/105/4960902>.
[61] Kevin Davis and Florencia Marotta-Wurgler, 'Contracting for Personal Data' (2019) 94 New York University Law Review 662.
[62] Parker and others (n 21).
[63] Ardion D Beldad, Menno De Jong and Michaël F Steehouder, 'When the Bureaucrat Promises to Safeguard Your Online Privacy: Dissecting the Contents of Privacy Statements on Dutch Municipal Websites' (2009) 26 Government Information Quarterly 559 <https://linkinghub.elsevier.com/retrieve/pii/S0740624X09000835>.

A long-standing strand of work, developed primarily by computer science scholars, concerns readability of privacy policies, that is, how easily it can be understood by an average reader given the fact that the legal and technical terms can be challenging to comprehend.[66] Becher and Benoliel,[67] for instance, run the Flesch-Kincaid readability tests (including the Flesch Reading-Ease and the Flesch-Kincaid Grade Level test) upon inspected 300 most popular websites and conclude that 96.76% of the examined privacy policies received a score lower than the recommended score of 60. The readability frame draws our focus closely upon the nature and quality of text, whereas the human, cognitive aspect of reading a privacy policy is equally important. It raises the issue of literacy, preference and capacity of the reader, asking whether special care and support are needed for vulnerable groups such as the elderly, children, the disabled. This latter aspect is moderately engaged in the empirical work, for instance, by Milkaite and Lievens who found the privacy policies of three most children-favoured services – Instagram, Snapchat & TikTok – are not suitable for them to read, calling for legal visualisation, co-design/co-creation techniques, and participatory design methods.[68]

Apart from the implications for legal crafting and design, natural language processing technologies are utilised to simplify privacy policies and render them easy-to-read for all. Automated tools have been developed to summarise lengthy and lawyer-written texts, check the completeness of a sample policy, or identify non-compliance. Torre et al. report a precision rate of 85% (and recall rate of 96%) over an AI-assisted inspection of 24 privacy policies.[69] Tesfay et al. present an algorithm based on Naïve Bayes classifier capable of summarising lengthy privacy policies into short and condensed notes.[70] A more tailored approach is undertaken by Chang et al. that models on an individual's privacy preference before automatically extracting and analysing a given policy, with an accuracy rate of 81%.[71] The growing capacity to classify texts, highlight critical elements, identify contextual changes via automated means brings hope to reduce the level of textual complexity structurally.[72] Yet, the efficacy and effectiveness of such tools remain to be evaluated. As Sun and Xue observe, these products tend to be inadequate, incomplete, or characterised by an inclination to allow for excessive processing.[73]

---

[64] Federal Trade Commission, 'Privacy Online: Fair Information Practices in the Electronic Marketplace' (2000) <https://www.ftc.gov/sites/default/files/documents/reports/privacy-online-fair-information-practices-electronic-marketplace-federal-trade-commission-report/privacy2000.pdf>.

[65] Davis and Marotta-Wurgler (n 61).

[66] Barbara Krumay and Jennifer Klar, 'Readability of Privacy Policies' in Anoop Singhal and Jaideep Vaidya (eds), *Annual Conference on Data and Applications Security and Privacy* (Springer International Publishing 2020) <https://link.springer.com/content/pdf/10.1007%2F978-3-030-49669-2_22.pdf>.

[67] Shmuel I Becher and Uri Benoliel, 'Law in Books and Law in Action: The Readability of Privacy Policies and the GDPR', *Consumer Law and Economics* (Springer International Publishing 2021) <http://link.springer.com/10.1007/978-3-030-49028-7_9>.

[68] Ingrida Milkaite and Eva Lievens, 'Child-Friendly Transparency of Data Processing in the EU: From Legal Requirements to Platform Policies' (2020) 14 Journal of Children and Media 5.

[69] Damiano Torre and others, 'An AI-Assisted Approach for Checking the Completeness of Privacy Policies against GDPR', *Proceedings of the IEEE International Conference on Requirements Engineering* (2020) <https://orbilu.uni.lu/bitstream/10993/43548/1/An AI-assisted Approach for Checking the Completeness of Privacy Policies Against GDPR.pdf>.

Consent is one of the most troubled area of data protection law and has been subject to extensive doctrinal debates.[74] The empirical viewpoint could inform and add to these debates, with nuanced enforcement landscapes both before and after the GDPR. Borghi et al.'s inspection of 200 most popular UK-based e-commerce websites before the GDPR shows 'a significant departure between EU legal theory and practice'.[75] At that time, the notice-and-consent mechanism has evolved into a phase where the interfaces for eliciting consent are mostly created by the websites (69%), but mostly in a superficial way, with a small number of the websites genuinely seeking valid consent (16.2%).[76] Moreover, the willingness to be tracked (30.5%) or expressed refusal of consent (50%) are usually not respected, indicating sub-optimal level of implementation.[77] Burghardt et al. inspected a sample of 100 platforms operating in Germany, and a significant number of the platforms engaged neither provided adequate and clear information in their privacy policies nor asked users for consent.[78] Despite rapid response to the researchers' request for information or deletion of personal data, the information provided was insufficient.

The GDPR represents a critical opportunity that the EU brings clarity and granularity to its notice and consent mechanism. Ironically, the situation has not been changed much. For example, Dabrowski et al. evaluate the potential changes brought by the GDPR to cookie setting behaviours;

---

[70] Welderufael B Tesfay and others, 'I Read but Don't Agree: Privacy Policy Benchmarking Using Machine Learning and the EU GDPR', *Companion of the World Wide Web Conference,* (International World Wide Web Conferences Steering Committee 2018) <https://doi.org/10.1145/3184558.3186969>.
[71] Cheng Chang and others, 'Automated and Personalized Privacy Policy Extraction Under GDPR Consideration' in Edoardo S Biagioni, Yao Zheng and Siyao Cheng (eds), *Wireless Algorithms, Systems, and Applications* (Springer International Publishing 2019) <https://link.springer.com/content/pdf/10.1007%2F978-3-030-23597-0_4.pdf>.
[72] ibid.
[73] Ruoxi Sun and Minhui Xue, 'Quality Assessment of Online Automated Privacy Policy Generators', *Proceedings of the Evaluation and Assessment in Software Engineering* (ACM 2020) <https://dl.acm.org/doi/10.1145/3383219.3383247>.
[74] Solon Barocas and Helen Nissenbaum, 'Big Data's End Run around Procedural Privacy Protections' (2014) 57 Communications of the ACM 31 <https://dl.acm.org/doi/10.1145/2668897>; Elettra Bietti, 'Consent as a Free Pass: Platform Power and the Limits of the Informational Turn' (2020) 40 Pace Law Review 306 <https://core.ac.uk/download/pdf/286538264.pdf>; Julie Cohen, 'How (Not) to Write a Privacy Law' [2021] Knight First Amendment Institute at Columbia University <https://knightcolumbia.org/content/how-not-to-write-a-privacy-law>; Viljoen (n 11); Salomé Viljoen, Jake Goldenfein and Lee McGuigan, 'Design Choices: Mechanism Design and Platform Capitalism' (2021) 8 Big Data & Society 205395172110343 <http://journals.sagepub.com/doi/10.1177/20539517211034312>.
[75] M Borghi, F Ferretti and S Karapapa, 'Online Data Processing Consent under EU Law: A Theoretical Framework and Empirical Evidence from the UK' (2013) 21 International Journal of Law and Information Technology 109 <https://academic.oup.com/ijlit/article-lookup/doi/10.1093/ijlit/eat001>.
[76] ibid.
[77] ibid.
[78] Thorben Burghardt and others, 'A Study on the Lack of Enforcement of Data Protection Acts', *International Conference on e-Democracy* (2010) <http://link.springer.com/10.1007/978-3-642-11631-5_1>.

their inspection of top 100,000 websites on Alexa shows notable improvements in the EU jurisdictions but not others, thus concluding with a 'two-class internet'.[79] For the EU users, the encounters with unconditional usage of persistent cookies reduced significantly, with 49% websites (of top 1000 websites) and 26% websites (of top 100,000 websites) refraining from placing cookies without consent.[80] Before some US states began to borrow the GDPR, it also exerted an extraterritorial impact on US users as well, albeit to a lessened degree, with the overall cookie load reduced by 46.7%.[81] In a similar vein, Degeling et al.'s measurement of 500 most popular websites selected for each EU member state (6579 in total) indicates an increase to 84.5% in the percentage of websites publishing a privacy policy, although the increase halted abruptly after the GDPR took effect.[82] They also documented a 16% increase (compared to January 2018) in the number of websites in Europe (62.1%) displaying consent pop-ups or interfaces.[83] Yet, Kollnig et al.'s interrogation of third-party tracking shows that only 4% of the market players had established a sufficient lawful basis under the GDPR.[84] It can be gleaned from these patterns that explicitly non-compliant patterns such as lacking the means to either reject or revoke are gradually fading away.[85] Even if a complaint mechanism of consent is put in place, in reality, decision-making process might be adversely affected by how the interface is designed.

Yet, the question quickly shifts from the absence of tools or interfaces to express consent to one concerning the quality, usefulness, and potential presence of deceptiveness or mislead in those interfaces. Related empirical efforts have captured this shift, examining the willingness and capacity of users to make autonomous choices and revealing the impacts of these technological interfaces. These efforts focus on examining the effectiveness of these mechanisms and the ways in which interface design can influence user behaviour and consent. Interestingly, this cluster of evidence in part inspired by the GDPR enforcement has not resulted in a reform of the GDPR itself, but somewhat considered and reflected in the EU's new Digital Services Act that explicitly restricts dark patterns, specific interfaces designed to trick or mislead the process of giving consent and other decision-making processes.[86] The term dark pattern, coined by human-computer interaction scholar Harry Brignull[87] and ostensibly inspired by behavioural economist Richard Thaler's work on nudging,[88] refers to user interface that typically exploits the position, colour, font size, among many other aspects of an interface to trick, deceive and mislead users, thereby exerting a deliberate impact on user decision-making in the processing of their personal data. In Gray et

---

[79] Adrian Dabrowski and others, 'Measuring Cookies and Web Privacy in a Post-GDPR World', *International Conference on Passive and Active Network Measurement* (2019) <http://link.springer.com/10.1007/978-3-030-15986-3_17>.
[80] ibid.
[81] ibid.
[82] Martin Degeling and others, 'We Value Your Privacy ... Now Take Some Cookies: Measuring the GDPR's Impact on Web Privacy' (2019) 42 Informatik Spektrum 345 <http://link.springer.com/10.1007/s00287-019-01201-1>.
[83] ibid.
[84] Konrad Kollnig and others, 'A Fait Accompli? An Empirical Study into the Absence of Consent to Third-Party Tracking in Android Apps A Fait Accompli? An Empirical Study into the Abs', *Proceedings of the Seventeenth Symposium on Usable Privacy and Security* (2021) <https://www.usenix.org/conference/soups2021/presentation/kollnig>.
[85] Utz and others (n 20).
[86] Article 25, Digital Services Act.
[87] Harry Brignull, 'Dark Patterns' (*Deceptive Patterns*, 2010) <https://www.deceptive.design/>.
[88] Richard H Thaler and Cass R Sunstein, *Nudge: Improving Decisions About Health, Wealth and Happiness* (Yale University Press 2009).

al.'s terms, dark patterns are 'instances where designers use their knowledge of human behaviour (e.g., psychology) and the desires of end users to implement deceptive functionality that is not in the user's best interest'.[89]

Around the time the GDPR takes effect, a body of empirical work is dedicated to scrutinising dark patterns in consent interfaces (or "pop-ups"), working towards nuanced definitions, taxonomies, and depiction of a specialised industry. Initial empirical investigations into dark patterns tend to map and categorise dark patterns while estimating its current scale. Marthur et al., for instance, conduct a first examination at scale (~53K product pages from ~11k shopping websites), identifying around 11.1% of the websites utilising text-based dark patterns and classifying them in 15 types and 7 broad categories.[90] The automated approach undertaken restricts the scope of their examination to text-based user interfaces, and the total number of dark patterns would be higher if visual, audio and video types of interfaces are inspected. In contrast, Soe et al. manually inspects 300 consent notices from the news outlets, concluding with 297 out of 300 websites using dark patterns when eliciting consent from their users.[91] Nouwens et al. scraped 10,000 most popular websites in the UK using the designs of five most popular consent management system or platforms (CMPs).[92] As the research shows, implicit consent (32.5%) and pre-ticked options (30.3%-56.2%) are fairly common, 'reject all' button generally scarce (50.1% do not have one), and only 11.8% meet the minimal requirements.[93] Some configurations do have a measurable impact on user decision-making, with the removal of opt-out button from the first page increasing 22-23% of consent giving, and more granular control on the first page decreasing 8-20%. In this regard, Utz et al. present more comprehensive experiments with over 80,000 users to measure the influence of dark patterns on user decision-making.[94] Their research emphasises relatively high impacts of lower-left part of the screen, binary choice, and nudging through highlighting and pre-selection on user decision-making. Similarly, working with 150 University students from Germany and Austria, Machuletz and Böhme's experiment evaluates the impact of the number of options (choice proliferation) and the highlighted default button (nudging).[95] As they conclude, choice proliferation has a real impact on participants making regrettable choices whereas nudging's effect on decision-making and perception is negligible. The effect of nudging designs is also evaluated by Fernandez et al.'s research on interactions made by 1100 MTurk workers.[96] According to them, colour-based nudging bars are reported to have a stronger influence than purpose-stated interfaces on the participants' behaviour.

---

[89] Colin M Gray and others, 'The Dark (Patterns) Side of UX Design', *Proceedings of the 2018 CHI Conference on Human Factors in Computing Systems* (ACM 2018) <https://dl.acm.org/doi/10.1145/3173574.3174108>.
[90] Arunesh Mathur and others, 'Dark Patterns at Scale: Findings from a Crawl of 11K Shopping Websites' (2019) 3 Proceedings of the ACM on Human-Computer Interaction.
[91] Than Htut Soe and others, 'Circumvention by Design - Dark Patterns in Cookie Consent for Online News Outlets', *Proceedings of the 11th Nordic Conference on Human-Computer Interaction: Shaping Experiences, Shaping Society* (ACM 2020) <https://dl.acm.org/doi/10.1145/3419249.3420132>.
[92] Nouwens and others (n 20).
[93] ibid.
[94] Utz and others (n 20).
[95] Machuletz and Böhme (n 20).
[96] Carlos Bermejo Fernandez and others, 'This Website Uses Nudging: MTurk Workers' Behaviour on Cookie Consent Notices', *Proceedings of the ACM on Human-Computer Interaction* (2021) <https://dl.acm.org/doi/10.1145/3476087>.

Cookie interfaces are part of CMPs in rapid development. A specialised industry emerges to offer solutions with which users can express their preference in a convenient, rapid and machine-readable manner. The problem with CMPs, however, lies in its inscrutability, i.e., that individuals are unable to check whether their expressed preferences are rightly documented, respected and acted upon. Matte et al.'s study on IAB's Europe's Transparency and Consent Framework, which crawls logs of how 22,949 European websites store and process user consent, reveals 141 cases where positive consent was registered even if users have not made such, 232 cases of pre-selecting boxes, and 54% (out of 560 inspected websites) having failed to meet the GDPR requirements.[97] Empirical insights show the actual level of compliance by these industrial solutions. Further, they are also instrumental in ascertaining legal matters, such as what legal roles do providers of consent management platforms play. Santos et al.'s deep-dive into two major providers reveals that CMPs do process personal data and on some occasions act as controllers, contrary to common characterisation.[98] Lastly, research using online surveys to collect and analyse qualitative data about dark patterns is relatively scarce. Kulyk et al.'s survey with 150 participants shows that many participants tend to view cookie disclaimers as 'nuisance' offering little or no aids in decision-making and that the participants tend to judge by website reputation and service types.[99]

Data subject rights have been a popular area for empirical investigation and, compared to dark patterns, this area of research is methodologically more challenging due to the need for human interaction as part of the evidence-gathering. In the absence of community-level consensus on methodology, divergences exist in request strategies, sample size, orienting questions, and normative reflections.[100] At the early phase of GDPR enforcement, data subject rights have been primarily inspected by researchers who initiate GDPR requests (as data subjects) to a sample of companies, which we categorise as a form of experiment (instead of content analysis). Sometimes, the GDPR requests often engage more than one right, i.e., the joint use of access and portability rights to expand the scope of access,[101] or that of portability and erasure rights to ensure that one could ideally leave a service with all the personal data taken away.[102] A common pattern of early-stage empirical research is that they are mostly oriented towards preparedness or compliance,[103] with exceptions that place data subject rights in broader contexts e.g. trust[104] or research.[105] Partly due to this compliance-based approach, not all data subject rights have been equally and significantly examined from an empirical point of view, and our review shows a great disparity in the extent to which such rights are empirically engaged.

---

[97] Célestin Matte, Nataliia Bielova and Cristiana Santos, 'Do Cookie Banners Respect My Choice? Measuring Legal Compliance of Banners from IAB Europe's Transparency and Consent Framework' [2019] arxiv <http://arxiv.org/abs/1911.09964>.

[98] Cristiana Santos and others, 'Consent Management Platforms Under the GDPR: Processors and/or Controllers?', *Annual Privacy Forum: Privacy Technologies and Policy* (2021) <https://link.springer.com/10.1007/978-3-030-76663-4_3>.

[99] Oksana Kulyk and others, '"This Website Uses Cookies": Users' Perceptions and Reactions to the Cookie Disclaimer', *Proceedings 3rd European Workshop on Usable Security* (Internet Society 2018) <https://www.ndss-symposium.org/wp-content/uploads/2018/06/eurousec2018_12_Kulyk_paper.pdf>.

[100] Mika Raento, 'The Data Subject's Right of Access and to Be Informed in Finland: An Experimental Study' (2006) 14 International Journal of Law and Information Technology 390 <https://academic.oup.com/ijlit/article/14/3/390/789294> accessed 5 January 2022; Jan Tolsdorf, Michael Fischer and Luigi Lo Iacono, 'A Case Study on the Implementation of the Right of Access in Privacy Dashboards' in Nils Gruschka and others (eds), *Privacy Technologies and Policy* (Springer International Publishing 2021).

[101] Tobias Urban and others, 'A Study on Subject Data Access in Online Advertising After the GDPR', *International Workshop on Cryptocurrencies and Blockchain Technology* (2019).

[102] Dominik Herrmann and Jens Lindemann, 'Obtaining Personal Data and Asking for Erasure: Do App Vendors and Website Owners Honour Your Privacy Rights?' [2016] Lecture Notes in Informatics (LNI) <https://arxiv.org/abs/1602.01804v2>.

[103] Urban and others (n 101).

The right of access has solicited significant empirical attention in various contexts, including surveillance,[106] work,[107] research,[108] litigation,[109] and law enforcement (security).[110] It has been exercised against a wide variety of controllers and/or processors e.g. (immigration) authorities,[111] app vendors,[112] online websites (with privacy dashboard),[113] digital advertising platforms.[114] The right is leveraged not just for knowing about the processing of personal data, the original intention of this right, but utilised for genomic sequencing access[115] or for evidence-based legal reform.[116] Despite the diverse contexts in which this right has been applied, the empirical landscape features two seemingly conflicting narratives. On the one hand, the right appears rife with uncertainties and challenges that prevent effective and meaningful exercise in reality.[117] The notion that the right is too individualistic or atomic by nature, with limited or no structural effects, has been empirically

---

and theoretically contested.[118] On the other hand, the right is susceptible to abuse as an impersonation tool, and it has been characterised in the security literature as a distinctive form of attack. Di Martino et al., for instance, reveal the implications of poor authentication processes in the context of the right of access.[119] Despite that the concern is genuine, it gives incentive to controllers to be more demanding in the authentication processes, reportedly creating significant hurdles for effective and meaningful exercise of the right. The line between secure response to access requests and one of unduly obstructive remains elusive.

The right to erasure, alternatively known as the 'right to be forgotten', is empirically inspected in two main aspects. First, self-reported data are collected from various actors about the right's perceived performance. As revealed by Mangini et al.'s two surveys, respectively of 385 individual users and 43 executive employees, organisations tend to point out their struggling with compliance, indicative of a perceived positive impact on privacy.[120] Second, empirical attention has been given to the design characteristics that adversely affect a user's ability to navigate and make the right choice. Habib et al., for instance, provide an extensive content analysis of 150 English-language websites.[121] Their results show that, despite the offering of privacy choices on websites, the choice dashboards are inconsistently located, unusable due to missing or unhelpful information, and disingenuous due to the links that did not lead to the stated choice.[122] Focus groups are organised in another research by Habib et al. to detect the level of ease for average internet users to exercise deletion, with practical hurdles identified as preventing average individuals in fulfilling their intentions.[123] There are also cognitive challenges for what deletion really means. Murillo et al.'s study with 22 participants and 7 experts reveals that deletion has several technical variations by which the personal data are processed in one way or another, e.g., to remove from an account, server, or access from others, or temporal disappearance.[124] Hence, there is little or no way of verifying the right-based request of deletion without clarity on the technical status and outcomes.

The right to data portability (RtDP) is subject to extensive empirical inspection due to its apparent technical and complex nature. Attempts have been undertaken to understand its recognition and explanation in privacy policies on the one hand, and how it is instructed and exercised in practice on the other. This right has more prominent relevance on the internet of things (IoT) industry than others due to the need for continuous and seamless access and exchange of data, and multiple empirical research focus on the right's exercisability in this particular sector. Turner et al., for instance, examine 160 privacy policies provided by IoT producers, showing a low degree of compliance maturity.[125] Half of the policies never mentioned data portability or interoperability in any capacity, and none ever described the process in which the right might be transferred directly from one system to another. Turner et al also engage four widely available systems, on which they find that the RtDP was not yet exercisable in the IoT environments.[126] Further, the study by Barth has a larger sample (with 34 IoT platforms engaged) and concludes similarly with an unsatisfactory level of implementation.[127] As it reveals, 'big players' have more professional and faster dealing with requests than start-ups due to the existence of dedicated departments on security and data protection. An even larger sample collected by Wong and Henderson – 320 GDPR requests – reveals the disparities in the interpretation and compliance with the legal requirements concerning data format, i.e., the provision of personal data being structured, commonly used and machine-readable.[128] Wong and Henderson (2019) further argue that clarity to the standard of 'commonly used' could be brought by independent third-party assessment, and to that of 'machine-readability' by legal harmonisation on the concept of machine-readability.[129] Apart from its technical complexity, the right to data portability has been disputed for a lack of meaningful use cases. Yet, Zwiebelmann and Henderson empirically consider the right as an audit tool operationalised at an individual user level in the context of fitness tracker.[130] The right can be useful in understanding, revealing and challenging inaccuracies of data held by controllers, but practical challenges arise from incomplete provision of data, insufficient descriptions and metadata, and how data are interpreted and represented.

---

[118] René LP Mahieu, Hadi Asghari and Michel Van Eeten, 'Collectively Exercising the Right of Access: Individual Effort, Societal Effect' (2018) 7 Internet Policy Review.
[119] Mariano Di Martino and others, 'Personal Information Leakage by Abusing the GDPR "Right of Access"', *Proceedings of the 15th Symposium on Usable Privacy and Security, SOUPS 2019* (USENIX Association 2019) <https://www.usenix.org/conference/soups2019/presentation/dimartino>.
[120] Vincenzo Mangini, Irina Tal and Arghir-Nicolae Moldovan, 'An Empirical Study on the Impact of GDPR and Right to Be Forgotten - Organisations and Users Perspective', *Proceedings of the 15th International Conference on Availability, Reliability and Security* (ACM 2020) <https://dl.acm.org/doi/10.1145/3407023.3407080>.
[121] Hana Habib and others, 'An Empirical Analysis of Data Deletion and Opt-Out Choices on 150 Websites', *Proceedings of the 15th Symposium on Usable Privacy and Security* (USENIX Association 2019) <https://www.usenix.org/conference/soups2019/presentation/habib>.
[122] ibid.
[123] Hana Habib and others, '"It's a Scavenger Hunt": Usability of Websites' Opt-Out and Data Deletion Choices', *Proceedings of the 2020 CHI Conference on Human Factors in Computing Systems* (ACM 2020) <https://dl.acm.org/doi/10.1145/3313831.3376511>.
[124] Ambar Murillo and others, 'If I Press Delete, It's Gone": User Understanding of Online Data Deletion and Expiration', *Fourteenth Symposium on Usable Privacy and Security* (USENIX Association 2018).

Ostensibly the most doctrinally contested and technically complex right, the right not to be subject to automated decision-making is moderately reflected in the empirical scholarship, largely oriented towards, and inspired by, doctrinal debates on the so-called 'right to explanation'.[131] Serveto (2020)'s engagement with 43 news recommender systems reveals a particular concern about the lack of knowledge and compliance readiness about the right to explanation, in addition to the tactic of automating the provision of standardised answers.[132] Whereas 21% of the providers had not provided non-automated answers after two months of interaction, 42% responded with non-automated answers within 19 days, and another 28% provided their responses between 30 and 59 days, which exceeds the standard period required by law (i.e., one month). Moreover, 56% of the cases required reminders to be sent for a non-automated answer to be provided, and 13% required more than four reminders. The engagement by Dexe et al. with seven insurance companies (representing 90-95% of the Swedish home insurance market) shows great disparities in the ways in which requests for 'meaningful information' were responded, with three out of seven companies having not provided any information about the basic workings of the insurance, and another three responses unduly delayed without a legitimate reason.[133] It is on Art 15 GDPR, rather than Art 22, that Dexe et al.'s requests were made, meaning the premium might not be conducted on the basis of profiling or automated decision-making.[134] Yet, the 'right to explanation' nonetheless can be understood as resting in part upon Art 15 GDPR.[135]

---

[125] Turner and others (n 37).
[126] ibid.
[127] Marlene Barth, 'A Case Study on Data Portability' (2021) 45 Datenschutz und Datensicherheit - DuD 190 <https://doi.org/10.1007/s11623-021-1416-3>.
[128] Janis Wong and Tristan Henderson, 'How Portable Is Portable?: Exercising the GDPR's Right to Data Portability', *Proceedings of the 2018 ACM International Joint Conference and 2018 International Symposium on Pervasive and Ubiquitous Computing and Wearable Computers* (ACM 2018) <https://dl.acm.org/doi/10.1145/3267305.3274152>.
[129] Janis Wong and Tristan Henderson, 'The Right to Data Portability in Practice: Exploring the Implications of the Technologically Neutral GDPR' (2019) 9 International Data Privacy Law 173 <https://doi.org/10.1093/idpl/ipz008>.
[130] Zoe Zwiebelmann and Tristan Henderson, 'Data Portability as a Tool for Audit', *Adjunct Proceedings of the 2021 ACM International Joint Conference on Pervasive and Ubiquitous Computing and Proceedings of the 2021 ACM International Symposium on Wearable Computers* (Association for Computing Machinery 2021) <https://doi.org/10.1145/3460418.3479343>.
[131] For example: Sandra Wachter, Brent Mittelstadt and Luciano Floridi, 'Why a Right to Explanation of Automated Decision-Making Does Not Exist in the General Data Protection Regulation' (2017) 7 International Data Privacy Law 76; Andrew D Selbst and Julia Powles, 'Meaningful Information and the Right to Explanation' (2017) 7 International Data Privacy Law 233 <http://academic.oup.com/idpl/article/7/4/233/4762325>.
[132] Maria Mitjans Serveto, 'Exercising GDPR Data Subjects' Rights: Empirical Research on the Right to Explanation of News Recommender Systems' (2020) 6 European Data Protection Law Review (EDPL) <https://heinonline.org/HOL/Page?handle=hein.journals/edpl6&id=625&div=&collection=> accessed 21 January 2022.

## 3. Impact

The third cluster of evidence concerns the broader marked effects of the GDPR, intended and unintended, regardless of the legislative objectives. When it comes to the impact of the GDPR, little can or has been done to empirically measure the extent to which the law's stated objectives are met. Rather, existing work measures primarily the economic and competitive effects, in contrast to teleological or functional evaluation that refers strictly to the provisions or to the stated objectives. From a doctrinal point of view, the GDPR has its distinctive normative pursuits other than consumer welfare and competition, despite the potential convergences with other realms of law.[136] Yet, this is where the empirical work makes a unique contribution by revealing the desired or unintended consequences.

The evaluation of economic and competitive effects of the GDPR has been primarily undertaken in a particular realm (e.g., financial services), sector, or about a specific type of controller (e.g., multi-national organisations). For instance, Walczuch and Steeghs's qualitative data, derived from interviews with specialists from 23 multi-national organisations based in the Netherlands and

---

[133] Jacob Dexe, Jonas Ledendal and Ulrik Franke, 'An Empirical Investigation of the Right to Explanation Under GDPR in Insurance', *Trust, Privacy and Security in Digital Business: 17th International Conference* (Springer 2020) <https://portal.research.lu.se/en/publications/an-empirical-investigation-of-the-right-to-explanation-under-gdpr> accessed 21 January 2022.
[134] ibid.
[135] Wachter, Mittelstadt and Floridi (n 131).
[136] Albertina Albors-Llorens, 'Competition and Consumer Law in the European Union: Evolution and Convergence' (2014) 33 Yearbook of European Law 163 <https://academic.oup.com/yel/article-lookup/doi/10.1093/yel/yeu001>; Wolfgang Kerber, 'Digital Markets, Data, and Privacy: Competition Law, Consumer Law and Data Protection' (2016) 11 Journal of Intellectual Property Law & Practice 856 <https://academic.oup.com/jiplp/article-lookup/doi/10.1093/jiplp/jpw150>; Francisco Costa-Cabral and Orla Lynskey, 'Family Ties: The Intersection between Data Protection and Competition in EU Law' (2017) 54 Common Market Law Review 11 <https://kluwerlawonline.com/journalarticle/Common+Market+Law+Review/54.1/COLA2017002>; Natali Helberger, Frederik Zuiderveen Borgesius and Agustin Reyna, 'The Perfect Match? A Closer Look at the Relationship between EU Consumer Law and Data Protection Law' (2017) 54 Common Market Law Review 1427.

Germany, present perceived difficulties in transferring customer data across boundaries.[137] As they observe, the specialists did not consider the Directive as a non-tariff barrier to trade at the eve of the Directive's implementation, and compliance was believed by some to have a positive effect on business due to the enhanced trust from the customers. Xuereb et al.'s semi-structured interviews with 63 GDPR experts echo some of the previous observations, e.g., that compliance helps improve the trust, standardisation and reputation of the institutions they represent.[138] Their research is intended to evaluate the GDPR's impact (reduced to compliance economic costs) in European countries with less than 3 million population (Malta, Slovenia, Luxembourg, Lithuania, Latvia, Estonia, and Cyprus) but with general findings about increased workload and compliance costs.[139] In their view, such costs have a greater impact on underdeveloped European countries, and the principle of proportionality is thus emphasised for the GDPR enforcement.[140] Building upon 3.3 million survey responses from those who had been randomly exposed to 9,596 online banner advertising campaigns, a large-scale study predating the GDPR focuses on advertising effectiveness. It shows that the laws do affect advertising effectiveness by reducing the effectiveness of banner ads by 65% on average (in terms of changing stated purchase intent).[141] As the study concludes, this would ultimately affect the direction of innovation on advertising-supported internet but not necessarily alter the structure of current digital advertising ecosystem. A more concerning conclusion lies at the uneven distribution of the effectiveness loss: websites that offered general content (e.g., news and web services), with smaller ad presence, and those with no additional interactive, video or audio ad features suffered greater loss than those infused primarily with commercial content, thereby exerting an indirect impact on journalism and other undertakings primarily funded by advertising. A loss-loss situation is marked in the implementation of the law in digital advertising industry that requires reflections upon the configuration of the digital advertising industry and the law's underpinning and/or refining role, not limited to what Goldfarb & Tucker explicitly argued about the trade-off between consumer privacy and benefits brought by the ad-supported internet.

The economic and competitive effect of the GDPR stands as another arena for empirical evaluation. Goldberg et al., for instance, reveal a 12% reduction in Adobe's website pageviews and e-commerce revenues after the GDPR takes effect.[142] Another measurement by the same research group concludes with a 15% reduction in the use of web technology vendors in Europe by 27,000 top international websites.[143] Market concentration has been pinned as a direct and unintended result of GDPR enforcement.[144] As Johnson et al. reveal, vendor usage reduction is concomitant with 17% increase in vendor market concentration, with the beneficiaries being primarily those specialising in pervasive tracking such as Facebook and Google.[145] As such, the position of SMEs is increasingly disadvantaged. Peukert et al. (2022) coined the term 'regulatory spillovers' to contend that businesses are somewhat affected by the GDPR but not on equal terms, with tech giants enlarging their market shares in advertising and analytics, and SMEs having a worsening position to compete accordingly.[146]

---

[137] Rita M Walczuch and Lizette Steeghs, 'Implications of the New EU Directive on Data Protection for Multinational Corporations' (2001) 14 Information Technology & People 142 <https://www.emerald.com/insight/content/doi/10.1108/09593840110695730/full/html>.

[138] Kieran Xuereb and others, 'The Impact of the General Data Protection Regulation on the Financial Services' Industry of Small European States' (2019) VII International Journal of Economics and Business Administration 243 <http://ijeba.com/journal/342>.

[139] ibid.

[140] ibid.

[141] Avi Goldfarb and Catherine E Tucker, 'Privacy Regulation and Online Advertising' (2011) 57 Management Science 57 <http://pubsonline.informs.org/doi/10.1287/mnsc.1100.1246>; Catherine Tucker, 'Empirical Research on the Economic Effects of Privacy Regulation' (2012) 10 Journal on Telecommunications & High Technology Law 265.

[142] Samuel Goldberg, Garrett Johnson and Scott Shriver, 'Regulating Privacy Online: An Economic Evaluation of the GDPR' (2024) 16 American Economic Journal: Economic Policy 325.

[143] Garrett Johnson, Scott Shriver and Samuel Goldberg, 'Privacy & Market Concentration: Intended & Unintended Consequences of the GDPR' [2023] Management Science 1 <https://www.ssrn.com/abstract=3477686>.

The SME perspective is also qualitatively examined, revealing a handful of misconceptions and misorientations that prevent SMEs from bringing benefits or actual changes to market structure. For instance, Härting et al.'s work on the GDPR's impact on SEMs consists of 13 interviews with German data protection experts, intended to identify perceived influences of the new GDPR on SMEs (hypotheses), and 103 survey responses completed by German-speaking experts, designed to test the hypotheses developed.[147] With six primary factors identified (including costs, provision of information, process adaption, know-how, uncertainty, and expenditure of time), they conclude that all determinants have a general negative influence, but the former four appear relatively significant. Sirur et al.'s semi-structured interviews with 12 partitioners suggest that compliance seems 'reasonable and doable' for large organisations, giving skilled individuals the freedom of interpretation, but the same 'risk[s] alienate smaller companies who did not have the resources necessary to cope with the resulting overhead'.[148] Further, Norval et al.'s qualitative probe on 15 UK-based tech start-ups shows that their approaches to GDPR compliance are often misoriented, and support is needed from authorities in the form of awareness-raising, guidance as well as innovative measures e.g., regulatory sandbox.[149] The increase in the scrutiny of SMEs might facilitate a race-to-the-top effect, driven by privacy enhancing innovations as well as responsible use of personal data.

## 4. Clarity

The GDPR is composed of a mix of principles, rules and methods. Evolving from a principle-based framework, the EU data protection law sees an improved level of textual granularity by way of establishing the GDPR. Yet, there lie some concepts and principles under-interpreted due to the scarcity of guidelines or case-law, thereby inviting perspectives from various disciplines including law. In our sample, a specific cluster of evidence is dedicated to collecting views and expertise on a particular mechanism of the GDPR in order to improve its conceptual or regulatory clarity.

---

[144] Michal S Gal and Oshrit Aviv, 'The Competitive Effects of the GDPR' (2020) 16 Journal of Competition Law & Economics 349 <https://academic.oup.com/jcle/article/16/3/349/5837809>.
[145] Goldberg, Johnson and Shriver (n 142).
[146] Christian Peukert and others, 'Regulatory Spillovers and Data Governance: Evidence from the GDPR' (2022) 41 Marketing Science 746 <http://pubsonline.informs.org/doi/10.1287/mksc.2021.1339>.
[147] Ralf-Christian Härting, Raphael Kaim and Dennis Ruch, 'Impacts of the Implementation of the General Data Protection Regulations (GDPR) in SME Business Models—An Empirical Study with a Quantitative Design', *Agents and Multi-Agent Systems: Technologies and Applications* (2020) <http://link.springer.com/10.1007/978-981-15-5764-4_27>; Ralf Christian Härting and others, 'Impacts of the New General Data Protection Regulation for Small- and Medium-Sized Enterprises', *Proceedings of Fifth International Congress on Information and Communication Technology* (2021) <http://link.springer.com/10.1007/978-981-15-5856-6_23>.
[148] Sean Sirur, Jason RC Nurse and Helena Webb, 'Are We There yet? Understanding the Challenges Faced in Complying with the General Data Protection Regulation (GDPR)' [2018] Proceedings of the ACM Conference on Computer and Communications Security 88.
[149] Chris Norval and others, 'Data Protection and Tech Startups: The Need for Attention, Support, and Scrutiny' (2021) 13 Policy & Internet 278 <https://onlinelibrary.wiley.com/doi/10.1002/poi3.255>.

Accountability stands as a foundational principle of the GDPR but despite its wide breadth and conceptual richness, it has been passably engaged. Nadine and Nico organise four exploratory, in-depth interviews with Data Protection Officers (DPOs) from large Swiss organisations, coupled with two-phased qualitative content analysis, to synthesise their hands-on experiences about compliance.[150] Specifically, factors that might influence DPOs' behaviour with an impact on organisational privacy are examined, with two primary hurdles are identified, including "not being in a vital control function" with implications for their independence and authority, and the impracticability and ambiguity of the law that pose practical challenges. Based on 24 qualitative interviews with 29 enterprise architects, Huth et al. inquire whether and how Enterprise Architecture Management (EAM) may serve as a basis for the work of data protection management.[151] The results confirm their hypothesis that EAM does contribute, despite not being perfectly aligned, to compliance. In the context of personalisation, Biega et al. bring clarity to the principles of data minimisation and purpose limitation as applied to recommendation algorithms. In the absence of a homogenous definition, they propose two performance-based interpretations of data minimisation (global average algorithm performance versus local per-user minimum performance) while evaluating their effects.[152] Nyman and Große empirically consider security concerns from an IT-consulting viewpoint.[153] Based on 6 individual semi-structured interviews and 47 survey responses (response rate 58.75%), they identify challenges of various sorts that IT consulting firms encounter, including typical ones (e.g., weak awareness, inadequate experiences and ineffective communications) and new ones brought by the GDPR and related regulations (e.g., correct interpretation, timely reporting and adaption of server-level agreements and policies).[154]

There are also concepts newly injected into the EU data protection law, the implications of which remain ambiguous and contested. Based on a survey of 253 practitioners from China's IT industry, Bu et al. evaluate the factors that may influence "privacy by design" (PbD) implementation at both individual and organisational levels.[155] Their research shows that, although the workload is increased as a result, both individual decision-making and organisational control do influence whether and the extent to which PbD is implemented in a particular organisation.[156] On the same topic, Hadar et al. conduct in-depth interviews with 27 software developers of different backgrounds, and their grounded analysis shows interplay between several factors jointly affecting developers' decision-making, including favoured vocabulary of security (rather than privacy), organisational privacy climate, as well as software architectural patterns.[157] Rommetveit et al. present disparities in how PbD is interpreted, understood and implemented, among a 'techno-epistemic network' of professionals, identifying tensions and contradictions between various approaches.[158]

---

[150] Casutt Nadine and Ebert Nico, 'Data Protection Officers: Figureheads of Privacy or Merely Decoration?', *Proceedings of the 16th European Conference on Management Leadership and Governance* (ACPI 2020) <https://www.academic-conferences.org/wp-content/uploads/dlm_uploads/2020/10/000_ECMLG-2020-abstract-booklet-with-cover.pdf#page=37>.

[151] Dominik Huth and others, 'Empirical Results on the Collaboration between Enterprise Architecture and Data Protection Management during the Implementation of the GDPR', *Proceedings of the 53rd Hawaii International Conference on System Sciences* (2020).

[152] Biega and others (n 19).

[153] Maja Nyman and Christine Große, 'Are You Ready When It Counts? IT Consulting Firm's Information Security Incident Management', *Proceedings of the 5th International Conference on Information Systems Security and Privacy* (SCITEPRESS - Science and Technology Publications 2019) <http://www.scitepress.org/DigitalLibrary/Link.aspx?doi=10.5220/0007247500260037>.

[154] ibid.

[155] Fei Bu and others, '"Privacy by Design" Implementation: Information System Engineers' Perspective' (2020) 53 International Journal of Information Management 102124 <https://doi.org/10.1016/j.ijinfomgt.2020.102124>.

[156] ibid.

## IV. Discussion

The findings of our systematic review provide a complex empirical landscape of the EU data protection law, underlining the necessity for a more nuanced understanding of the GDPR's effectiveness, particularly in an evaluative context. It becomes clear that the regulation's effects extend beyond straightforward compliance actions for the EU data protection law is not a coherent set of rules with each of its components coordinated and synchronised. Rather, as our synthesis shows, it is a mix of heterogenous rules that jointly shape and achieve the broad imperatives of data protection, with broader socio-economic impacts, anticipated or unintended.

Our synthesis hence pinpoints a crucial reality that the unique characteristics of the GDPR render the evaluative attempts rather challenging. To conduct a reliable and adequate evaluation one must acknowledge the law's complex macro- and micro-structure, its diverse goals, the intricate and carefully balanced legal framework it offers, and varied contexts in which it operates. Overall, we observe two main hurdles in which GDPR-related evidence supports an informed and adequate evaluation. On the one hand, the heterogeneous nature of the GDPR and lack of coordination between its components make a coherent and general-level evaluation largely challenging. On the other hand, different components are at different stages of development, and evidence about their operation and enforcement is far from consistent and solid.

It can also be gleaned from this synthesis that the common critiques of the GDPR concerning the lack of enforcement against big tech, inadequate fines imposed, as well as its design flaws (over-reliance on consent) are relevant, but they are also partial and may misorient evaluative work away from meaningful measurement. These hyped debates are an integral part of the evaluation, yet it requires a careful thinking of the anticipated effects of the GDPR – e.g. whether it is expected to bring in structural remedies (as are epitomised by record-breaking fines) as other more relevant and specific legislative frameworks (e.g. EU competition law and more recent pro-competition rules in the Digital Markets Act are in place) In our view, while erected for the first time with hefty fines in the EU legal order, the GDPR takes its effects not by command-and-control approach, as research on the risk-based approach indicates,[159] but more of a meta regulation that foster a privacy-friendly nature.[160] It serves more to establish a privacy infrastructure on which individual remedies, organisational accountability as well as regulatory scrutiny can be operated than to providing market cures. As such, its effectiveness should be much more evaluated by the extent to which compliance mechanisms are put in place in an effective manner and which individual requests and interactions can be reliably and effectively responded. Following this risk-based approach, its maturity should be assessed by the level of support given to the start-ups as much as how big tech are penalised for non-compliance. Another instance is dark patterns, the legality and controversy of which broke out mostly in the context of GDPR but has increasingly been discussed in the EU legal order within other regulatory contexts such as consumer law reform or the new Digital Services Act. This reflects the complex and intricate and evolving interplay between the web of rules within the EU legal order, with the nature and position of the GDPR constantly evolving and adapting.

---

[157] Irit Hadar and others, 'Privacy by Designers: Software Developers' Privacy Mindset' (2018) 23 Empirical Software Engineering 259.
[158] K Rommetveit, A Tanas and N van Dijk, 'Data Protection by Design: Promises and Perils in Crossing the Rubicon Between Law and Engineering', *Privacy and Identity Management: The Smart Revolution* (2018) <https://link.springer.com/10.1007/978-3-319-92925-5_3>.
[159] Raphaël Gellert, *The Risk-Based Approach to Data Protection* (Oxford University Press 2020) <https://global.oup.com/academic/product/the-risk-based-approach-to-data-protection-9780198837718?cc=gb&lang=en&#> accessed 29 January 2021.

The internal contradiction between different types of mechanisms fabricated into the law is flagged in the doctrinal scholarship[161] as well as policy discourses[162] but less acknowledged in the formation and design of empirical evaluation. Our synthesis has brought it back to the surface and urgently requires deliberation in the empirical context. A more operational approach would take a level of granularity down further to look at the effectiveness of the GDPR's components, such as consent, transparency, data subject rights, in association with their defined objectives and anticipated effects as interpreted by the court and authoritative guidance. At this level, our synthesis serves as a solid basis for such specific-level evaluation, and the existing evaluative work should pay due attention to the evidence produced at the component level.

This synthesis further raises the critical question of whether the GDPR can be evaluated with reference to its overarching objectives, i.e. the protection of personal data, which has been substantiate diversely in practice as safeguarding the effective exercise of data subject rights, accountable data processing activities, robust and proactive enforcement, among many others. We argue that any evaluation of the GDPR requires a nuanced, multi-level framework that considers not only the achievement of its objectives properly defined and substantiated in specific contexts but consider all of its components and their interplay with other normative values and legal interplays. Different set of metrics should be developed for different types of tools, whether they be technical, organisational or conceptual. To inform public policy debates, it is also crucial to reflect on the GDPR's impacts on businesses, research, public interest undertakings that are not necessarily reflected within its stated objectives.

In any evaluative endeavour, a trickier issue concerns the dealing with a strand of empirical work produced mostly by economists and/or industry experts about economic and industrial impact of the GDPR, often represented in a negative manner as hindering innovation and economic growth. For a law that gives expression to a fundamental right within a legal order, it might be reasonable to argue that interests such as consumer welfare and competitiveness are related but should not be considered as admissible for evaluation. Achieving a high level of data protection might require the taking of economic and social costs that this strand of empirical evidence reveals. We contend that careful dealing with this cluster of evidence is important, particularly for empirical evidence that is evidently politicised or lacking rigour. In the meantime, there are also reliable but understated revelations on the unintended consequences of the GDPR that one shall not turn a blind eye for evaluation. The side effect or spill-over effects of the regulation has minimally been engaged, often as a reactive response to such political acts branded as research. It is crucial that a comprehensive evaluation attends not only to the validity and effects of GDPR mechanisms, but also its implications for values not recognised as legislative objectives, as well as its interplay with other related adjacent legal fields, the interplay between which are constantly evolving in the legal EU order as new laws such as the Digital Services Act, Digital Markets Act, Data Act and Artificial Intelligence Act are conceived on top of, and inevitably in interaction with the GDPR. Against this backdrop, the exact meaning and pursuit of data protection should be recalibrated against the evolving legal landscape and the interpretation of the protection of personal data in various settings.

---

[160] Colin Bennett and Charles Raab, *The Governance of Privacy: Policy Instruments in Global Perspective* (MIT Press 2006).
[161] Veale, Binns and Ausloos (n 60).
[162] European Data Protection Supervisor, 'Opinion 7/2015 Meeting the Challenges of Big Data A Call for Transparency, User Control, Data Protection by Design and Accountability' (2015) <https://www.edps.europa.eu/sites/default/files/publication/15-11-19_big_data_en.pdf>.

Lastly, our synthesis underlies the urgent need to coordinate and guide empirical efforts arising from different fields of study and prompted by diverse needs (academic, industrial or policymaking). We make our best effort to weave these empirical efforts into a structured and consistent narrative, but it is far from coordinated and organised in a matter that can inform policy-making or doctrinal debates. A general lack of agreed-upon methodology or research agenda is a major concern, and we are equally concerned with the concentration of efforts in realms with easy-to-acquire data, leaving various institutional, procedural and other organisational aspects in need of empirical inspection less attended. Our sample shows that, due to the ease of collection, the empirical landscape of the GDPR is pre-dominated by self-reported data on the one hand, and publicly available data on the other.[163] The over-representation of perceptive data or publicly available data significantly shapes the orientation and focus of the current empirical research agenda, potentially skewing findings towards more accessible but less nuanced insights. This may hinder a comprehensive understanding of the GDPR's deeper impacts and obscure complex dynamics that require more rigorous, methodologically diverse approaches to data collection and analysis. It also contributes also to uneven distribution of empirical attention and efforts across 99 GDPR provisions, centralising resources and attention on topics about which data can be gathered easily, with other novel and/or critical ones minimally engaged.

---

[163] See figure 5

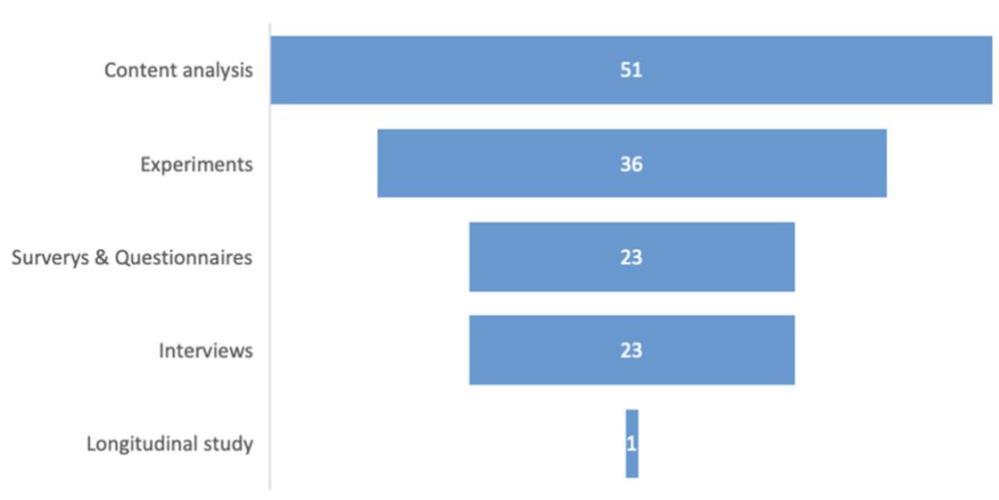
Figure 5 Distribution of research methods

Such a disproportional concentration is counterproductive for evidence-based evaluation, and coordination and guidance are urgently needed. Further, the lack of connection between empirical scholarship and the commissioned surveys or other statistical work is also a source of concern. Our sample shows that empirical data produced for evaluative work tend to focus on perception and other subjective metrics whereas the critical empirical data on the nuances and dynamics of compliance at organisational or local level cannot find its way to support and shape policymaking. We call for a collective endeavour to collect, pool and regularly update evidence about the GDPR while making useful connections to evaluative undertakings of various sorts. The establishment of a register or database of empirical evidence to monitor, document and integrate new evidence generated over time may be a good starting point. More fundamentally, consensus should be facilitated on many fronts to work towards a community-level research agenda, including the limitations of self-reported data and publicly scrapable data, selection of methods, and research resource distribution.

## V.     Conclusion

In this paper, we inspect, synthesise, and analyse research of an empirical nature produced over a timespan of nearly three decades (1995-March 2022) on EU data protection laws. Our motivation for synthesising this empirical evidence stems from the sobering reality that most of the doctrinal and legal work often lacks a nuanced understanding of how the GDPR operates in practice. For many members of the data protection communities, despite high interest in knowing what's happening in practice, empirical evidence in this domain is only sporadically referenced due to a lack of synthesis and meaningful discussion. Overall, there appears to be an evident undervaluation of the relevance and significance of empirical evidence when assessing the impact, objectives, and the delicate balance between data protection and other normative values and legal frameworks.

Our synthesis stands as an antithesis to the popular but problematic orientations towards lopsided understanding and evaluation of the EU data protection law. It invites a critique of the prevalent success/failure framework commonly seen in public discourses and exerting an influence over a part of our reviewed articles, setting hypothetical questions, and potentially leading to an over-simplistic, reduced understanding of the GPDR that overlooks the foundational changes the law has brought to individuals, society, as well as the privacy culture. A simplistic success/failure framework is fundamentally flawed as it presupposes that the law is internally coherent and consistent, with each of its mechanisms well aligned and mutually reinforcing. This fails to acknowledge the inherent complexity and occasional discord between various components of the GDPR, as our synthesis reveals, which can lead to conflicts and inefficiencies rather than a seamless regulatory environment. Further, our synthesis also reflects current focus on big tech and hefty fines imposed upon them. It may be justifiable to place regulatory priorities on major players due to their extensive data collection and processing activities, yet this focus can skew our understanding of the GDPR's overall effectiveness. Indeed, the imposition of fines on tech giants is often see as direct evidence of the GDPR's effectiveness, but this interpretation is partial and can be easily misled as it represents only one aspect of the GDPR's regulatory impact.

At the time of writing, proposals for streamlining or reforming the GDPR and related laws have been made at different levels. Whereas some member states of the EU seek to reduce the bureaucracies of the GDPR,[164] the UK's making radical moves redefining 'direction' of data protection by emphasising a pro-innovation approach.[165] Although surveys and statistical findings are provided as part of the legislative processes, the extent to which these reforms are evidence-based is open to discussion. We argue that evidence produced in two decades of empirical scholarship has a certain role in informing and shaping these processes, and endeavours urgently needed to systematise and validate the existing evidence, coupled with efforts to guide and coordinate related resources. Our synthesis of two decades of evidence provides a basis for this undertaking, but much more needs to be done at an institutional and cultural level to synchronise officially commissioned surveys or statistics with the scholarship to provide a more comprehensive empirical landscape and validate evidence not produced in a rigorous, peer-reviewed manner. For empirical research communities, more coordinated and community-level research agenda is required, directing empirical resources and efforts towards underappreciated aspects, mechanisms and methods. Effort should also be undertaken to establish meaningful connection to policymaking processes on the one hand, and doctrinal debates on the other. We assert that empirical evidence amassed over the past two decades should critically inform and shape policymaking, particularly through recent periodic evaluations, and that doctrinal debates would be significantly enriched by considering and reflecting upon these empirical findings. This paper, as a bridge between doctrinal undertakings and empirical inquiries, establishes the foundation for such critical endeavours.

---

[164] Ministry of Economics, Finance and Industrial and Digital Sovereignty, 'Reducing Bureaucracy in These Unprecedented Times - French-German Paper on Better Regulation and Modern Administration in Europe' (*Ministry of Economics, Finance and Industrial and Digital Sovereignty*, 2023) <https://presse.economie.gouv.fr/12102023-reducing-bureaucracy-in-these-unprecedented-times-french-german-paper-on-better-regulation-and-modern-administration-in-europe/>; OneTrust, 'Germany: BMF Proposes Reduction in Bureaucratic Requirements under Data Protection Laws' (*Data Guidance*, 2024) <https://www.dataguidance.com/news/germany-bmf-proposes-reduction-bureaucratic>.
[165] UK Department for Digital Culture Media & Sport (n 14).

# Appendix A: The codebook for the systematic review

| General scope | Empirical studies | Studies based on verifiable observation or experience rather than theory or logic, supported by empirical evidence, including first-hand primary data and official secondary data, qualitative and quantitative, which are collected as part of the article or report[166] |
|---|---|---|
| GDPR components examined in the reviewed paper | Public attitudes | The perception, understanding and expectation of the general public on the GDPR |
| | GDPR compliance | The overall level of compliance of data controllers and data processors with the GDPR |
| | Regulatory impacts | The impacts of enforcement of the GDPR on the markets and business |
| | Data protection principles | The principle as set out in Article 5 of the GDPR |
| | Consent | The ways of obtaining informed consent by data controllers as a legal basis for data processing and its validity according to Article 7 of the GDPR |
| | Transparency | Transparency obligations of data controllers as required by the GDPR, in particular, Articles 12-14, for example, in the form of privacy policies |
| | Data processor | The role of data processor and its processing activities according to Articles 28-29 of the GDPR |
| | Right of access | Right of access as required in Article 15 of the GDPR |
| | Right to rectification | Right to rectification as required in Article 16 of the GDPR |
| | Right to be forgotten | Right to be forgotten as required in Article 17 of the GDPR |
| | Right to restriction of processing | Right to restriction of processing as required in Article 18 of the GDPR |
| | Right to data portability | Right to data portability as required in Article 20 of the GDPR |
| | Right to object | Right to object as required in Article 21 of the GDPR |
| | Right not to be subject to automated decision-making | right not to be subject to automated decision-making as required in Article 22 of the GDPR |
| | Data protection by design and by default | Data protection by design and by default as required in Article 25 of the GDPR |
| | Security of data processing | The security of data processing as required in Article 32 of the GDPR |
| | Data protection impact assessment (DPIA) | DPIA and the prior consultation as required in Articles 35 and 36 of the GDPR |
| | Data protection officer (DPO) | The role of data protection officer in an organisation as required in Articles 37-39 of the GDPR |
| | Supervisory authorities | The functioning of data protection authorities in the Member States |

---

[166] The Copyright Evidence Wiki: Empirical Evidence for Copyright Policy. CREATe Centre: University of Glasgow (http://CopyrightEvidence.org)

|  | Penalties and liabilities | Penalties and liabilities resulting from GDPR violations issued by supervisory authorities |
|---|---|---|
|  | CJEU decision-making | GDPR-related judgements by the Court of Justice of the European Union |
| Methods | Longitudinal study | Continuous observation and repeated testing over prolonged periods of time |
|  | Interview | Structured, semi-structured or unstructured interviews carried out with stakeholders |
|  | Surveys & Questionnaires | Collection of qualitative or quantitative data through a survey or questionnaire |
|  | Experiments | Comparison of control and experimental groups |
|  | Content analysis | Analysis and interpretation of the text of documents to uncover the meanings and patterns |